\documentclass[iop,preprint,10pt]{emulateapj}

\usepackage{graphicx}
\usepackage{tabularx}
\usepackage{color}

\slugcomment{ }

\begin{document}
\title{Examining the secondary product origin of cosmic ray positrons with the latest AMS-02 data}

\author{Zhi-Qiu Huang\altaffilmark{1,2}, Ruo-Yu Liu\altaffilmark{1,2}, Jagdish C. Joshi\altaffilmark{1,2}, Xiang-Yu Wang\altaffilmark{1,2}}
\altaffiltext{1}{School of Astronomy and Space Science, Nanjing University, Nanjing 210093, China; ryliu@nju.edu.cn;
xywang@nju.edu.cn}
\altaffiltext{2}{Key laboratory of Modern Astronomy and Astrophysics (Nanjing University), Ministry of Education, Nanjing 210093, China}

\begin{abstract}
Measurements of cosmic-ray (CR) positron fraction by PAMELA and other experiments have found
an excess above 10  GeV relative to the standard predictions for secondary production in the interstellar medium (ISM). Although the excess has been mostly suggested to arise from some primary sources of positrons (such as pulsars and  or  annihilating dark matter particles), the almost constant flux ratio of $e^{+}/ \bar{p}$ argues for an alternative possibility that  the excess positrons and antiprotons up to the highest energies  are secondary products generated in hadronic interactions. Recently, \citet{2019PhRvD.100f3020Y} revisit this possibility by  assuming the presence of an additional population of CR nuclei sources. Here we examine this secondary product scenario using the \texttt{DRAGON} code, where the radiative loss of positrons is taken into account consistently.  We confirm that the CR proton spectrum and the antiproton data can be explained by assuming the presence of an additional population of CR  sources.
However, the corresponding positron spectrum deviates from the measured data significantly above 100~GeV due to the strong radiative cooling. This suggests that, although hadronic interactions can explain the antiproton data, the corresponding secondary positron flux is still not enough to account for the  AMS data. Hence contribution from some primary positron sources, such as pulsars or dark matter, is non-negligible.
\end{abstract}3


\section{introduction}
Inelastic interactions of primary cosmic rays (CRs) with the  interstellar medium are expected to produce secondary positrons and anti-protons with  quite soft spectra.
Unexpected spectra excess of positrons and antiprotons relative to the standard prediction has been detected at energies exceeding $E \gtrsim 10\,$GeV by PAMELA\citep{2009Natur.458..607A, 2012PhRvL.108a1103A}, the Alpha Magnetic Spectrometer(AMS-02)\citep{2015ICRC...34..300K,2019PhRvL.122d1102A} and other experiments.  The most popular model for the positron excess is that it arises from  some primary sources of positrons such as pulsars (e.g., \citep{2009JCAP...01..025H,2013PhRvD..88b3001Y,2012CEJPh..10....1P,2017JCAP...09..029J}), or  annihilating dark matter particles (e.g., \citep{2013PhRvD..88b3013C,2009PhRvD..79b3512Y}). However, measurements on the positron and antiproton flux by AMS-02 have revealed that the flux ratio $e^{+}/ \bar{p}$ keeps an almost constant value of $\simeq2$ in the energy range of 30-300 GeV, coincident with the expected ratio of $e^{+}/ \bar{p}$ produced as secondary particles from hadronic interactions of cosmic ray (CR) protons \citep{2017PhRvD..95f3009L, 2019arXiv190206173L}. Motivated by this coincidence, it is suggested that the excesses of positrons and antiprotons may result also from hadronic interactions \citep{2017PhRvD..95f3009L, 2019arXiv190206173L, 2010MNRAS.405.1458K, 2013PhRvL.111u1101B}. To explain the positron excess, \citet{2017PhRvD..95f3009L, 2019arXiv190206173L}  assume an energy independent diffusion coefficient and negligible energy losses up to the maximum energy observed (around TeV). Recently, \citet{2019PhRvD.100f3020Y} suggest that an additional population of primary CR nuclei sources beside the normal population of sources might have produced the excess secondary particles through the CR hadronic interactions.

However, positrons would suffer from efficient radiative loss in magnetic field and interstellar radiation field (ISRF) during their propagation in the interstellar medium (ISM), while anti-protons do not. Thus, the $e^{+}/ \bar{p}$ ratio in the propagated CR spectrum may not remain $\simeq 2$ at high energy as it does in the spectrum at generation. Recently, AMS-02 has released the latest measurement of CR positron data. The new measurement not only increases the statistics by a factor of three, but also extends the spectrum up to an unprecedentedly high energy, $\lesssim 1\,$TeV, and reveals a marked dropoff at the high-energy end \citep{2019PhRvL.122d1102A}. In the secondary origin model, one would naively expect a dropoff due to stronger radiative cooling for higher energy positrons. However, the exact position of the steepening due to radiative cooling depends on specific physical conditions, such as the magnetic fields in the galactic halo and the residence time of positrons. The precise measurement of AMS-02 on the positron spectrum motivates us to conduct a critical study of  whether the new positron data can be explained simultaneously with the antiproton data in the framework of the secondary origin model assuming the presence of an additional population of primary CR sources.

Two classes of the additional population sources of the primary CRs (i.e., the cosmic-ray accelerators) have been suggested. One is that  a single nearby supernova remnant (SNR) accelerates cosmic ray protons, which interact with a presumed dense gas cloud and produce secondary positrons and antiprotons (e.g., \citet{2009PhRvD..80f3003F, 2017PhRvD..96b3006L}). This model has limitations on the age and the distance of the source, i.e., the age of the SN should be around $10^5$ yr and the distance should be $\sim 100$ pc. However, evidence from deposition of $^{60}$Fe in the deep ocean crust suggests a SN occurred  2~Myr ago\citep{2002PhRvL..88h1101B, 1996ApJ...470.1227E, 2015ApJ...800...71F, 1999PhRvL..83...18K}, much older than the age required. The other one assumes a continuously distributed source population to produce secondary positrons and antiprotons \citep{2019PhRvD.100f3020Y}. With less constraints on the sources compared to the former model, the idea proposed by \citet{2019PhRvD.100f3020Y} seems to be more attractive and reasonable. Hence, we do not consider the single SNR scenario, but look into the latter scenario in which more free parameters are introduced for the second CR source population, providing a favourable condition for the reproduction of the excesses in the positron and antiproton spectra. The open code \texttt{DRAGON} \footnote{\url{https://github.com/cosmicrays/DRAGON}}  \citep{2010APh....34..274D,2013JCAP...03..036D,2017JCAP...02..015E} is introduced to solve the propagations of CRs and the production of secondary particles.

The paper is organized as follows. In \S 2, we first attempt to explain the CR proton data measured by DAMPE and other experiments using the \texttt{DRAGON} code. In \S 3, we calculate the secondary particle flux using \texttt{DRAGON} and  compared the results with the corresponding data. We present discussions in \S 4.

\section{model}
\subsection{Fitting the proton spectrum}
CRs spread in the galaxy through diffusive propagation after being injected from sources. In this work, we use \texttt{DRAGON} code to numerically solve the diffusive transport equation. The detailed description of \texttt{DRAGON} can be found in \citet{2008JCAP...10..018E} and the \texttt{DRAGON} website\footnote{\url{http://dragon.hepforge.org/.}}. With this code, CR nuclei and lepton propagation in the Galaxy can be successfully calculated. In our calculations, the halo is treated as a cylinder, with its radius $R_{\rm{max}} \,=\, 12$ kpc. Its half thickness, $L$, is adopted as 4 kpc. The diffusion coefficient $D$ is considered to be spatial independent and described as
\begin{equation}
D(\rho) = D_{0} \beta (\frac{\rho}{\rho_{0}})^{\delta},
\end{equation}
where $\beta$ is the CR speed in the unit of speed of light and $\rho$ is the rigidity.

Positrons and antiprotons are produced as secondaries by the hadronic interactions of the primary CRs (mainly protons). We start from the phenomenological fitting to the primary CR proton spectrum below 100\,TeV that is relevant for our study. Based on the data of AMS-02 \citep{2014PhRvL.113l1102A} and DAMPE \citep{2019arXiv190912860A}, we divide the proton spectrum into two components.  The first one is the conventional background cosmic-ray component (dubbed as the component 1), with the form of a single power law. The initial injection spectrum can be described as
\begin{equation}
N_{1}(\rho) \propto (\frac{\rho}{\rho_{0}}) ^{- \alpha_{0}}.
\end{equation}

Recent observations reveal that the proton spectrum hardens at around 200~GeV \citep{2011ApJ...728..122Y, 2019arXiv190912860A} and softens at $E$ $\simeq$ $10^4$ GeV\citep{2019arXiv190912860A}. The hardening of the proton spectrum requires a second, hard component,  which could originate from some extra sources (dubbed as the component 2), such as young massive stars \citep{2013MNRAS.429.2755B, 2019NatAs...3..561A}. We find that the initial injection spectrum of this second component can be written by a broken power law,
\begin{equation}
N_{2}(E)\propto
\left\{
\begin{array}{ll}
(\frac{\rho}{\rho_{1}}) ^{- \alpha_{1}}, & {\rm if}\,\, \rho < \rho_{1}, \\
(\frac{\rho}{\rho_{1}}) ^{- \alpha_{2}}, & {\rm if}\,\,\rho \geqslant \rho_{1},%
\end{array}%
\right.
\end{equation}

\begin{figure}
\begin{center}
\includegraphics[scale=0.37]{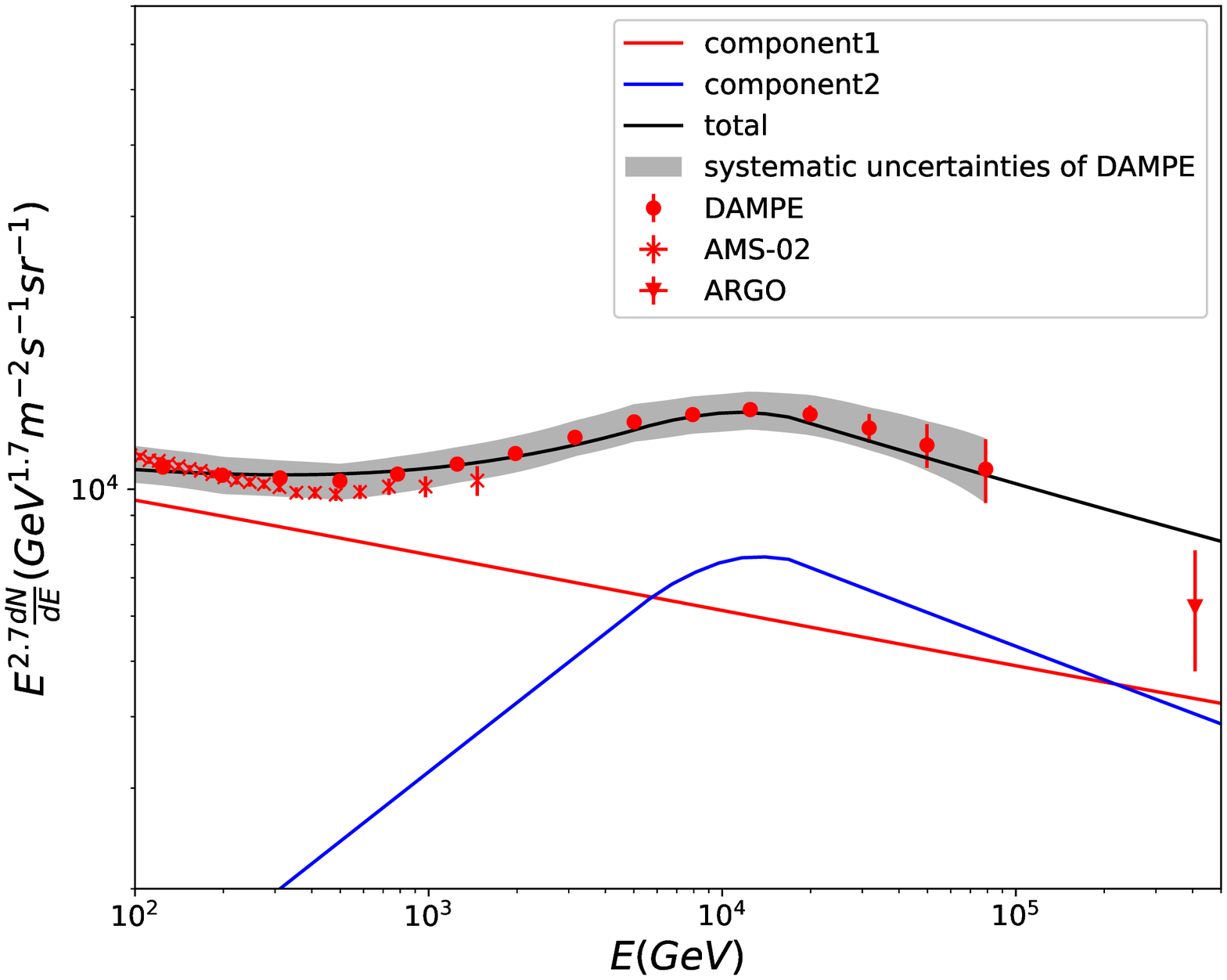}
\caption{Fit of the proton spectrum with two source populations of CRs . The red and blue lines represent the contributions from component 1 and component 2, respectively.  The total flux is shown by the black line. \label{fig:proton}}
\end{center}
\end{figure}

The respective contributions to proton flux from these two populations and the sum of them are shown in Fig.~\ref{fig:proton}, confronting with the measurement of AMS-02, DAMPE and ARGO experiments \citep{2014PhRvL.113l1102A, 2019arXiv190912860A,2011ASTRA...7...15A}. The adopted parameters are listed in Table \ref{tab:pd}.

\subsection{Calculation for the secondary particle flux}
During the propagation in the galaxy, primary CRs injected by two populations would interact with ISM and produce secondary particles. With \texttt{DRAGON} code, we can obtain these secondary CR fluxes when calculating the proton spectra simultaneously. Since these secondary particles are produced through hadronic interactions with ISM, we dub them as ISM-component 1 and  ISM-component 2, respectively.

In \citet{2019PhRvD.100f3020Y}, the leaky box model is used to calculate the secondary particle flux. They find that the secondary particles from the interaction between primary CRs and ISM (ISM-component 1 and ISM-component 2) are not enough to explain the antiproton data. To solve this problem, an extra grammage $X_{s}$ inside the second sources is introduced to produce additional antiparticles.

During the propagation of CRs, the magnetic field would scatter these particles in random motion with characteristic velocities relative to the Alfv\`{e}n speed ($v_{A}$), which causes a diffusion in momentum and results in a second-order Fermi acceleration.  This process is referred as "reacceleration" \citep{1995ApJ...441..209H} (see also \citet{2004APh....21...45B} for reacceleration via the first-order Fermi acceleration). Reacceleration can be taken into account in the \texttt{DRAGON} code. We find that the extra grammage is not necessary when the reacceleration is taken into account.
 This is because that when considering reacceleration, a smaller $\delta$ can be adopted to fit the B/C ratio. With a smaller $\delta$, the secondary particle spectrum becomes harder, and leads to more anti-protons above $\sim 100$\,GeV produced by primary protons during the propagation in the ISM.

On the premise of fitting both the proton flux and the antiproton flux, the simultaneously obtained positron flux, however, significantly deviates from the AMS-02 data, as will be discussed in the next section.

\section{results }
\begin{figure*}
\begin{center}
\includegraphics[scale=0.25]{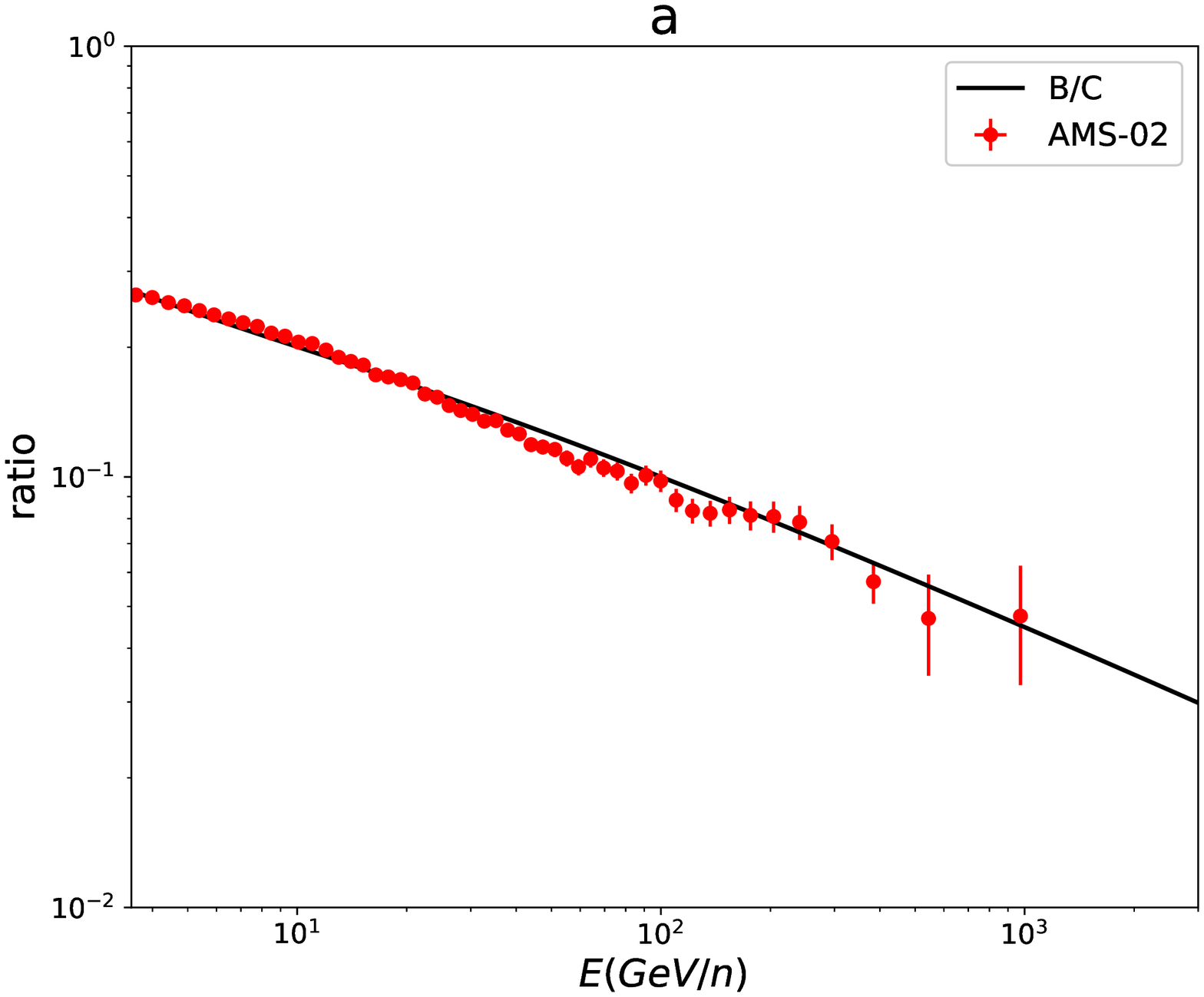}
\includegraphics[scale=0.25]{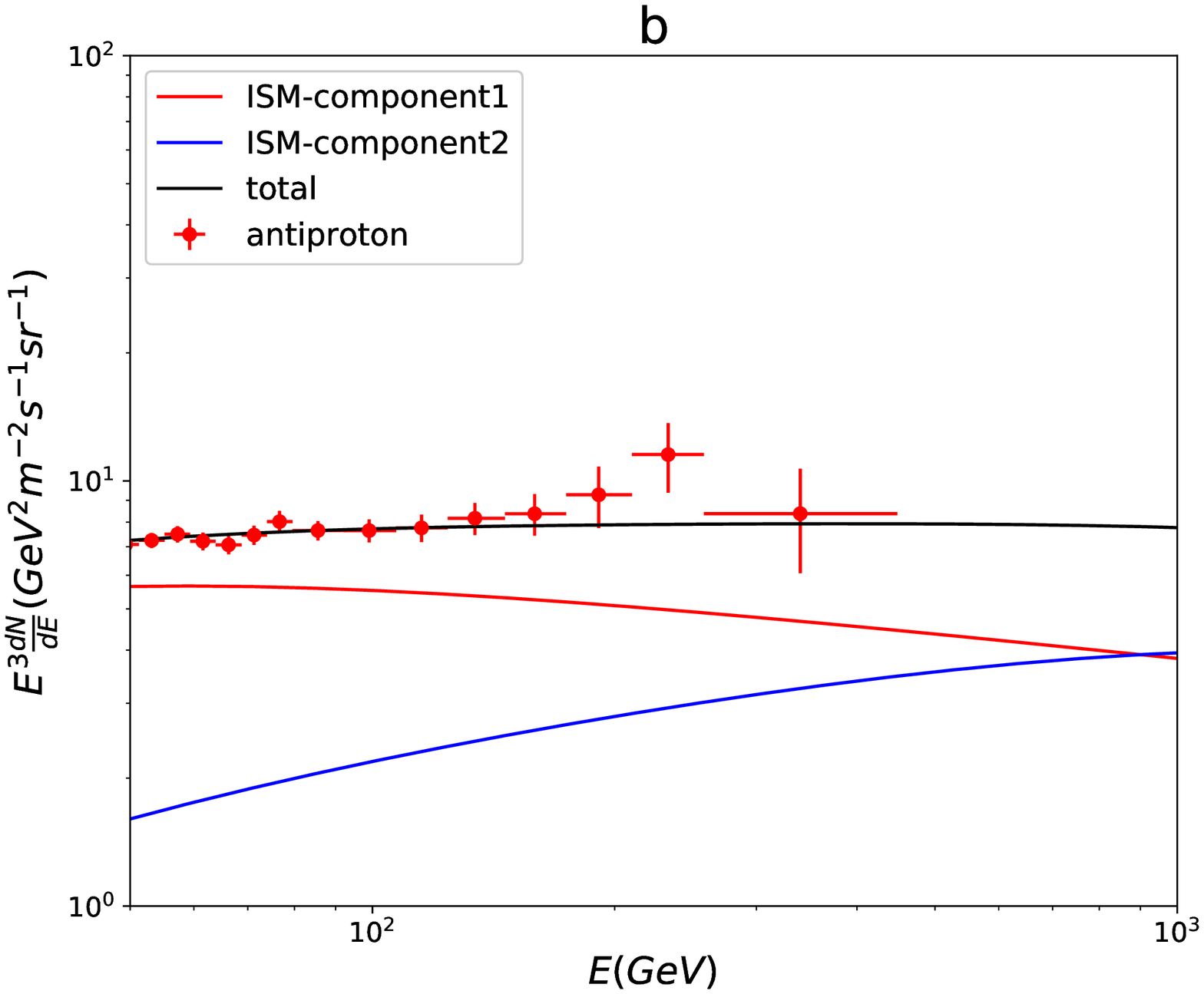}
\includegraphics[scale=0.25]{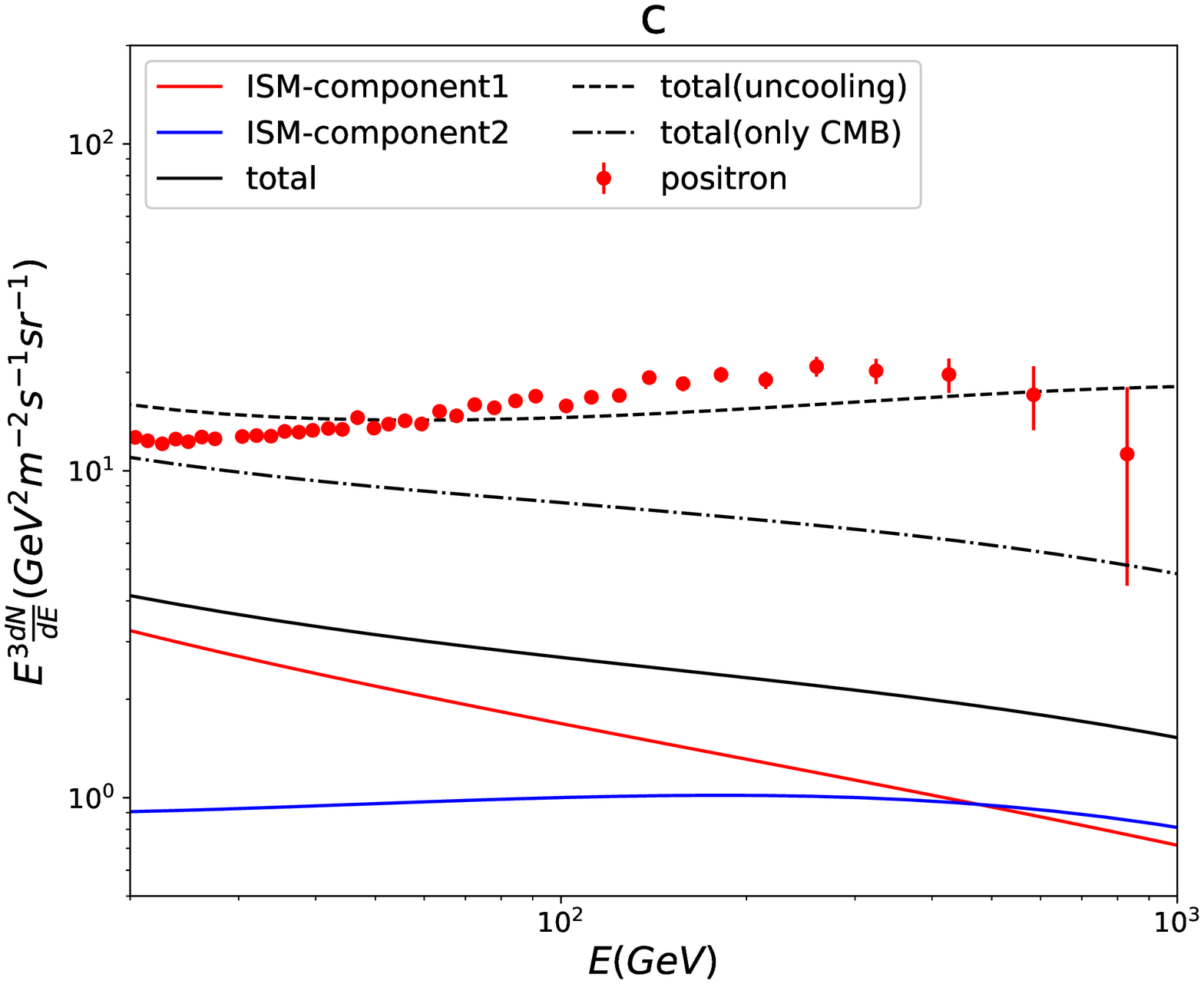}
\caption{Modelings of the B/C ratio (left panel), antiproton flux (middle panel) and positron flux (right panel). The data of B/C ratio are taken from \citet{2016PhRvL.117w1102A}. The data of antiproton and positron fluxes are taken from \citet{2016PhRvL.117i1103A} and \citet{2019PhRvL.122d1102A}, respectivly. The red and the blue lines represent the antiparticles produced from the ISM-component 1 and 2, respectively. The black solid line is the total flux. In the right panel, the black dashed line represents the total positron flux without considering the radiative cooling, and the black dashed-dotted line represents the total positron flux when only the inverse-Compton cooling due to CMB photons is considered.\label{fig:fitting}}
\end{center}
\end{figure*}

The resultant B/C ratio, antiproton flux and positron flux are shown in Fig.~\ref{fig:fitting}, 
as a function of energy per nucleon $E\, =\, \frac{Ze\rho}{A}$, where $e$ is the charge of a proton, $Z$ and $A$ are the atomic and mass number. For protons, antiprotons and positrons, $E\, =\, e\rho$. For other nuclei, $E\, \approx \, 0.5e\rho$.

The models and parameters we used are listed in Table \ref{tab:pd}. The spatial distributions for the two source components are considered to be the same.

\begin{table}
\caption{Model  parameter values for the propagation and the magnetic field used in the {\texttt{DRAGON}} code. In our calculation, we use the spatial distribution of two source populations provided by \citet{2001RvMP...73.1031F}, the gas distribution provided by \citet{2004A&A...422L..47S}, and the geometry of the galaxy magnetic field provided by \citet{2011ApJ...738..192P}. $A_{1}$ and $A_{2}$ are the normalization fluxes of ISM-component 1 and ISM-component 2 at 100~$\rm{GeV}$, respectively (in unit of $\rm{GeV}^{-1}\cdot \rm{m}^{-2}\cdot \rm{s}^{-1}\cdot \rm{sr}^{-1}$).\label{tab:pd}}
\begin{ruledtabular}
\begin{tabular}{ |p {4 cm}| l | }

Model/Parameter  &   Option/Value \\
 \hline
 Grid Type      &       2D\\
  $R_{\rm{max}}$													        &     12~kpc \\
  $L$                      														&     4~kpc  \\
  Gas Distribution         &      Galprop  \\
  Source Distribution          												&   Ferriere \\
  Diffusion type															&     Constant   \\
  $D_{0}$                               &    $4.3 \times 10^{28}$~$\rm{cm^{2} /s}$\\
  $\rho _{0}$                          &    4~GV\\
  $\delta$                  &   0.4\\
  $v_{A}$                  &   30~km/s\\
  Magnetic Field Type    &   Pshirkov \\
  $B_{0}^{\rm{disk}}$      &   $2.0 \times 10^{-6}$~Gauss \\
  $B_{0}^{\rm{halo}}$      &   $4.0 \times 10^{-6}$~Gauss \\
  $B_{0}^{\rm{turb}}$      &   $7.5 \times 10^{-6}$~Gauss \\
  $A_{1}$      &         $3.8 \times 10^{-2}$ \\
  $A_{2}$      &         $4.9 \times 10^{-3}$ \\
  $\alpha _{0}$        &    2.4\\
  $\alpha _{1}$        &    1.9\\
  $\alpha _{2}$        &    2.5\\
  $\rho _{1}$              &    12000~GV

\end{tabular}
\end{ruledtabular}
\end{table}

The B/C ratios of two source components are shown in Fig.~\ref{fig:fitting}(a), where the ratios from two components are the same. In Fig.~\ref{fig:fitting}(b), contributions to antiprotons from the ISM-component 1 and the ISM-component 2 are shown by the red and blue lines, respectively. Our results show that the antiproton data can be well explained by the sum of two components. Contributions to positrons by two source components are shown in Fig.~\ref{fig:fitting}(c). For  high-energy positrons, the radiative cooling due to the synchrotron radiation and inverse Compton scattering is important. As shown by the black solid line, the fluxes  drop dramatically starting from $\sim$ 10 GeV. To illustrate the importance of cooling effect on positrons, we also show the case without considering the cooling, denoted by the black dashed line in Fig.~\ref{fig:fitting}(c). Without considering any energy losses, the $e^{+}/ \bar{p}$ ratio is around 2, consistent with theoretical expectation \citep{2017PhRvD..95f3009L, 2019arXiv190206173L}. Since there are uncertainties in the strength of the magnetic field and the interstellar radiation field (ISRF), we also consider an extreme case for illustration, where only the cooling due to CMB photons are taken into account (see the black dashed-dotted line).  In such a case, the positron flux still cannot explain the AMS-02 data.

The above results can be understood by comparing the residence timescale of CRs, $\tau_{\rm res}$, with the radiative cooling timescale of positrons, $\tau_{\rm c}$. The former one can be derived from the so-called "grammage'', which measures the amount of material that  CRs collide with before they leave the Galaxy. The grammage  is defined by $X(E)={\bar n}m_{p}\tau_{\rm res}(E)c$, where ${\bar n}$ is the average density of gas in the Galaxy, $m_{p}$ is the mass of a proton, and $c$ is the speed of light.
The ratio between the secondary CR flux and their parent primary CR flux, such as the B/C ratio, is only sensitive to the grammage accumulated during the propagation. 
In the leaky-box approximation, this ratio can be written as \citep{2010MNRAS.405.1458K, 2019PhRvD.100f3020Y}
\begin{equation}\label{ratio}
R(E)=\frac {\frac {X(E)}{m_{p}} \sigma_{p \rightarrow s}} {1+\sigma _{t} \frac {X(E)}{m_{p}}}.
\end{equation}
Here $\sigma_{p \rightarrow s}$ is the differential cross section for the production of secondary particle, and $\sigma_{t}$ is the total inelastic cross section of a certain species of secondary particle. By fitting the B/C data, we obtain $X(E)\, =\, 2.1 (\frac{E}{300\, \rm{GeV}})^{-0.4}\, \rm{g/cm^{2}}$.

Then we can estimate the residence timescale given an average gas density of ${\bar n}=n_{\rm ISM}(l/L)$ \citep{2019arXiv190311584G}, where $n_{\rm ISM} \,\sim \, 1\, \rm{cm^{-3}}$ is the average density in the Galactic disk,  $l \,\sim \, 150$\,pc is the height of the disk, and $L$ is the height of the CR halo. Adopting a typical height of $L\, =\, 4$~kpc  and the grammage obtained above, we find the residence time of CRs is $\tau_{\rm {res}}\simeq 26\, {\rm Myr}$ at 300 GeV.



The main uncertainty in the above estimate lies in the height,  $L$, of the CR halo{\footnote{The CR halo is introduced to avoid producing too strong anisotropy of TeV CRs compared to the observations, as a larger $L$ can increase the residence timescale of CRs in the Galaxy and hence lead to a higher degree of isotropy \citep{2015APh....60...86E}. Moreover, there are other observational evidences supporting the existence of the halo \citep[e.g.,][]{2013MNRAS.436.2127O, 2010ApJ...724.1044S,2015ApJ...807..161T}.}}.
According to  previous studies, the typical height of CR halo is $3-10$\,kpc \citep[e.g.,][]{2013MNRAS.436.2127O, 2020PhRvD.101b3013E, 2001ICRC....5.1836M, 2011ApJ...729..106T}. Combining $\tau_{\rm{res}}(E)\,= \, L^{2}/D(E)$ with $X(E)= n_{\rm ISM}(l/L) m_p \tau_{res}(E) c$, we get $\tau_{\rm res}(E)\,=\, 26 \, (\frac{L}{4\, \rm{kpc}}) (\frac{E}{300 \, {\rm{GeV}}})^{-0.4}\,\rm{Myr}$ when the grammage is fixed. 

On the other hand, the cooling timescale of positrons, $\tau_{\rm c}$, can be given by
\begin{equation}
\tau_{\rm c}(E)\, = \, 2.5\,\, (\frac{300\, \rm GeV} {E})\, (\frac{1 \, {\rm{eV/cm^{3}}}}{U_{B}\, +\, U_{ph}})\,  \rm{Myr},
\end{equation}
where $E$ is the energy of positrons, $U_{B}$ is the energy density of the magnetic field and $U_{ph}$ is the energy density of the radiation field.
In the extreme assumption that only IC cooling due to the CMB radiation is considered, i.e.,  $U_{B}\, +\, U_{ph}\, =\, U_{\rm{CMB}}\, =\, 0.26 \, {\rm{eV/cm^{3}}}$, we obtain $\tau_{\rm c}(300 \, \rm{GeV})\, = \, 10$~Myr, which is already shorter than $\tau_{\rm {res}}(300 \, \rm{GeV})\, \gtrsim \, 20\, \rm{Myr}$. Therefore, any secondary-origin scenarios can not explain the positron data because of the severe radiative cooling.

Recent measurements  by AMS-02 shows that the spectral indices of secondary-to-primary ratio (e.g., Li/C, Be/C, B/C) exhibit a hardening of $0.13\, \pm \, 0.03$ above 200~GV \citep{2018PhRvL.120b1101A}. This may indicate that the slope ($\delta$) of the diffusion coefficient becomes smaller above 200 GV. As a smaller $\delta$ at high energy implies that the diffusion coefficient increases with energy more slowly, we obtain  a larger residence timescale ($\tau_{\rm res}\propto 1/D$), given that the residence timescale below 200 GV is fixed. That is, $\tau_{\rm res}(E)\, =\,  30 \, (\frac{L}{4\, \rm{kpc}}) (\frac{E}{300 \, {\rm{GeV}}})^{-0.27}\,\rm{Myr}$ for $E\, \geqslant \, 100 \, \rm{GeV}$, if $\tau_{\rm{res}}(E\, \leqslant \, 100\, \rm{GeV})$ is normalized to fit the B/C data.  Thus, the  cooling effect of positrons will be more significant in this case.


 Note that the scarcity of the positron flux above several tens of GeV cannot be compensated by introducing an additional positron component of secondary origin, since otherwise the co-produced antiproton flux will overshoot the data.
As a result, we conclude the secondary positron flux from the hadronic interactions is  insufficient to account for the  AMS-02 positron data when the spectrum of accompanying antiprotons  is tuned to match the data.

\section{Discussions and Conclusions}
In summary, by using the \texttt{DRAGON} code, we have shown that the CR proton data can be explained by two population of sources and  the hadronic interaction of these CR protons with ISM  can explain the antiproton data. However, the secondary positron flux simultaneously produced in this hadronic interaction model is insufficient to explain the measured data. This is due to the severe radiative energy losses of high-energy positrons, which demonstrates that some primary sources of positrons, such as dark matters and pulsars \citep[e.g.,][]{1995A&A...294L..41A, 1995PhRvD..52.3265A,2009JCAP...01..025H,2019arXiv190608542F}, are needed at higher energies to account for the additional positron fluxes.

We find that the extra grammage  introduced by \citet{2019PhRvD.100f3020Y} is not needed when the diffusive reacceleration is considered in the code.  There has been suggestion that  diffusive reacceleration is helpful in explaining the B/C ratio at low energy, given that 50\% of the total CR power is provided by reacceleration \citep{2017A&A...597A.117D}. However, whether the reacceleration really occurs at the required level is unknown (e.g., \citet{2019arXiv190311584G}), and further studies are needed to determine the contributions of reacceleration to the total CR power. On the other hand,  whether the reacceleration presents or not does not  change our conclusion about the secondary positron flux. As long as the antiproton flux matches data,  the positron flux is insufficient to account for the AMS-02 data due to the radiative cooling.

Some alternative secondary production scenarios, like the Nested Leaky Box (NLB) model \citep[e.g.,][]{2014ApJ...786..124C} do not need to introduce the CR halo. Instead, the model assumes an energy-independent diffusion coefficient or CR residence timescale above $\sim 200\,$GeV. As a result, TeV CRs can still have sufficient time to get isotropized before leaving Galaxy, and positrons of $\sim 100\,$GeV  can quickly leave Galaxy before cooling. However, the key assumption of the model, i.e., an energy-independent residence timescale above 200\,GeV, is inconsistent with the latest AMS-02 observation on the B/C ratio, as explicitly pointed out by \citet{2016PhRvL.117w1102A}.

The secondary positrons could also be accelerated in a subset of supernova remnant shocks which propagate into molecular clouds that are positively charged by cosmic-ray protons \citep{1990ICRC....4..109D, 2016PhRvD..94f3006M}. The acceleration could harden the positron spectrum. However, the maximum energy that this mechanism can work is limited to be less than 100~GeV according to the current understanding \citep{1990ICRC....4..109D, 2016PhRvD..94f3006M}.




\begin{thebibliography}{}

\bibitem[Ackermann et al.(2012)]{2012PhRvL.108a1103A} Ackermann, M., Ajello, M., Allafort, A., et al.\ 2012, \prl, 108, 011103

\bibitem[Adriani et al.(2009)]{2009Natur.458..607A} Adriani, O., Barbarino, G.~C., Bazilevskaya, G.~A., et al.\ 2009, \nat, 458, 607




\bibitem[Aguilar et al.(2014)]{2014PhRvL.113l1102A} Aguilar, M., Aisa, D., Alvino, A., et al.\ 2014, Physical Review Letters, 113, 121102




\bibitem[Aguilar et al.(2016)]{2016PhRvL.117i1103A} Aguilar, M., Ali Cavasonza, L., Alpat, B., et al.\ 2016, Physical Review Letters, 117, 091103


\bibitem[Aguilar et al.(2016)]{2016PhRvL.117w1102A} Aguilar, M., Ali Cavasonza, L., Ambrosi, G., et al.\ 2016, Physical Review Letters, 117, 231102


\bibitem[Aguilar et al.(2018)]{2018PhRvL.120b1101A} Aguilar, M., Ali Cavasonza, L., Ambrosi, G., et al.\ 2018, \prl, 120, 021101


\bibitem[Aguilar et al.(2019)]{2019PhRvL.122d1102A} Aguilar, M., Ali Cavasonza, L., Ambrosi, G., et al.\ 2019, Physical Review Letters, 122, 041102


\bibitem[Aharonian et al.(1995)]{1995A&A...294L..41A} Aharonian, F.~A., Atoyan, A.~M., \& Voelk, H.~J.\ 1995, \aap, 294, L41


\bibitem[Aharonian et al.(2019)]{2019NatAs...3..561A} Aharonian, F., Yang, R., \& de O{\~n}a Wilhelmi, E.\ 2019, Nature Astronomy, 3, 561


\bibitem[Amenomori et al.(2011)]{2011ASTRA...7...15A} Amenomori, M., Bi, X.~J., Chen, D., et al.\ 2011, Astrophysics and Space Sciences Transactions, 7, 15


\bibitem[An et al.(2019)]{2019arXiv190912860A} An, Q., Asfandiyarov, R., Azzarello, P., et al.\ 2019, arXiv e-prints, arXiv:1909.12860




\bibitem[Atoyan et al.(1995)]{1995PhRvD..52.3265A} Atoyan, A.~M., Aharonian, F.~A., \& V{\"o}lk, H.~J.\ 1995, \prd, 52, 3265




\bibitem[Ben{\'{\i}}tez et al.(2002)]{2002PhRvL..88h1101B} Ben{\'{\i}}tez, N., Ma{\'{\i}}z-Apell{\'a}niz, J., \& Canelles, M.\ 2002, Physical Review Letters, 88, 081101




\bibitem[Blasi(2004)]{2004APh....21...45B} Blasi, P.\ 2004, Astroparticle Physics, 21, 45


\bibitem[Blum et al.(2013)]{2013PhRvL.111u1101B} Blum, K., Katz, B., \& Waxman, E.\ 2013, \prl, 111, 211101




\bibitem[Bykov et al.(2013)]{2013MNRAS.429.2755B} Bykov, A.~M., Gladilin, P.~E., \& Osipov, S.~M.\ 2013, \mnras, 429, 2755


\bibitem[Cholis, \& Hooper(2013)]{2013PhRvD..88b3013C} Cholis, I., \& Hooper, D.\ 2013, \prd, 88, 023013


\bibitem[Cowsik et al.(2014)]{2014ApJ...786..124C} Cowsik, R., Burch, B., \& Madziwa-Nussinov, T.\ 2014, \apj, 786, 124




\bibitem[Di Bernardo et al.(2010)]{2010APh....34..274D} Di Bernardo, G., Evoli, C., Gaggero, D., et al.\ 2010, Astroparticle Physics, 34, 274


\bibitem[Di Bernardo et al.(2013)]{2013JCAP...03..036D} Di Bernardo, G., Evoli, C., Gaggero, D., et al.\ 2013, JCAP, 2013, 036


\bibitem[Dogiel \& Sharov(1990)]{1990ICRC....4..109D} Dogiel, A.~V., \& Sharov, S.~G.\ 1990, International Cosmic Ray Conference, 109


\bibitem[Drury, \& Strong(2017)]{2017A&A...597A.117D} Drury, L.~O. 'C ., \& Strong, A.~W.\ 2017, \aap, 597, A117


\bibitem[Ellis et al.(1996)]{1996ApJ...470.1227E} Ellis, J., Fields, B.~D., \& Schramm, D.~N.\ 1996, \apj, 470, 1227


\bibitem[Erlykin \& Wolfendale(2015)]{2015APh....60...86E} Erlykin, A.~D., \& Wolfendale, A.~W.\ 2015, Astroparticle Physics, 60, 86


\bibitem[Evoli et al.(2008)]{2008JCAP...10..018E} Evoli, C., Gaggero, D., Grasso, D., et al.\ 2008, JCAP, 2008, 018


\bibitem[Evoli et al.(2017)]{2017JCAP...02..015E} Evoli, C., Gaggero, D., Vittino, A., et al.\ 2017, JCAP, 2017, 015


\bibitem[Evoli et al.(2020)]{2020PhRvD.101b3013E} Evoli, C., Morlino, G., Blasi, P., et al.\ 2020, \prd, 101, 023013





\bibitem[Fang et al.(2019)]{2019arXiv190608542F} Fang, K., Bi, X.-J., \& Yin, P.-F.\ 2019, arXiv e-prints, arXiv:1906.08542


\bibitem[Ferri{\`e}re(2001)]{2001RvMP...73.1031F} Ferri{\`e}re, K.~M.\ 2001, Reviews of Modern Physics, 73, 1031


\bibitem[Fry et al.(2015)]{2015ApJ...800...71F} Fry, B.~J., Fields, B.~D., \& Ellis, J.~R.\ 2015, \apj, 800, 71


\bibitem[Fujita et al.(2009)]{2009PhRvD..80f3003F} Fujita, Y., Kohri, K., Yamazaki, R., \& Ioka, K.\ 2009, \prd, 80, 063003




\bibitem[Gabici et al.(2019)]{2019arXiv190311584G} Gabici, S., Evoli, C., Gaggero, D., et al.\ 2019, arXiv e-prints, arXiv:1903.11584




\bibitem[Heinbach \& Simon(1995)]{1995ApJ...441..209H} Heinbach, U., \& Simon, M.\ 1995, \apj, 441, 209



\bibitem[Hooper et al.(2009)]{2009JCAP...01..025H} Hooper, D., Blasi, P., \& Serpico, P.~D.\ 2009, JCAP, 2009, 025


\bibitem[Hooper et al.(2017)]{2017PhRvD..96j3013H} Hooper, D., Cholis, I., Linden, T., \& Fang, K.\ 2017, \prd, 96, 103013



\bibitem[Joshi, \& Razzaque(2017)]{2017JCAP...09..029J} Joshi, J.~C., \& Razzaque, S.\ 2017, JCAP, 2017, 029






\bibitem[Katz et al.(2010)]{2010MNRAS.405.1458K} Katz, B., Blum, K., Morag, J., \& Waxman, E.\ 2010, \mnras, 405, 1458


\bibitem[Knie et al.(1999)]{1999PhRvL..83...18K} Knie, K., Korschinek, G., Faestermann, T., et al.\ 1999, Physical Review Letters, 83, 18






\bibitem[Kounine(2015)]{2015ICRC...34..300K} Kounine, A.\ 2015, 34th International Cosmic Ray Conference (ICRC2015), 34, 300





\bibitem[Lipari(2017)]{2017PhRvD..95f3009L} Lipari, P.\ 2017, \prd, 95, 063009


\bibitem[Lipari(2019)]{2019arXiv190206173L} Lipari, P.\ 2019, arXiv:1902.06173


\bibitem[Liu et al.(2017)]{2017PhRvD..96b3006L} Liu, W., Bi, X.-J., Lin, S.-J., Wang, B.-B., \& Yin, P.-F.\ 2017, \prd, 96, 023006






\bibitem[Malkov et al.(2016)]{2016PhRvD..94f3006M} Malkov, M.~A., Diamond, P.~H., \& Sagdeev, R.~Z.\ 2016, \prd, 94, 063006


\bibitem[Moskalenko et al.(2001)]{2001ICRC....5.1836M} Moskalenko, I.~V., Mashnik, S.~G., \& Strong, A.~W.\ 2001, International Cosmic Ray Conference, 1836


\bibitem[Orlando \& Strong(2013)]{2013MNRAS.436.2127O} Orlando, E., \& Strong, A.\ 2013, \mnras, 436, 2127










\bibitem[Profumo(2012)]{2012CEJPh..10....1P} Profumo, S.\ 2012, Central European Journal of Physics, 10, 1


\bibitem[Pshirkov et al.(2011)]{2011ApJ...738..192P} Pshirkov, M.~S., Tinyakov, P.~G., Kronberg, P.~P., et al.\ 2011, \apj, 738, 192








\bibitem[Strong et al.(2004)]{2004A&A...422L..47S} Strong, A.~W., Moskalenko, I.~V., Reimer, O., et al.\ 2004, \aap, 422, L47


\bibitem[Su et al.(2010)]{2010ApJ...724.1044S} Su, M., Slatyer, T.~R., \& Finkbeiner, D.~P.\ 2010, \apj, 724, 1044


\bibitem[Tibaldo et al.(2015)]{2015ApJ...807..161T} Tibaldo, L., Digel, S.~W., Casandjian, J.~M., et al.\ 2015, \apj, 807, 161


\bibitem[Trotta et al.(2011)]{2011ApJ...729..106T} Trotta, R., J{\'o}hannesson, G., Moskalenko, I.~V., et al.\ 2011, \apj, 729, 106






\bibitem[Yang \& Aharonian(2019)]{2019PhRvD.100f3020Y} Yang, R., \& Aharonian, F.\ 2019, \prd, 100, 063020


\bibitem[Yin et al.(2009)]{2009PhRvD..79b3512Y} Yin, P.-F., Yuan, Q., Liu, J., et al.\ 2009, \prd, 79, 023512


\bibitem[Yin et al.(2013)]{2013PhRvD..88b3001Y} Yin, P.-F., Yu, Z.-H., Yuan, Q., et al.\ 2013, \prd, 88, 023001


\bibitem[Yoon et al.(2011)]{2011ApJ...728..122Y} Yoon, Y.~S., Ahn, H.~S., Allison, P.~S., et al.\ 2011, \apj, 728, 122





\end{thebibliography}
\end{document}